\documentclass[12pt,preprint]{aastex}

\newcommand{\simle}{\mbox{$\stackrel{<}{_{\sim}}$}}

\newcommand{\jmicron}{\,\hbox{$\mu$m} }

\slugcomment{To Appear in Astrophysical Journal Letters}

\shorttitle{WR 140 Proper Motions}
\shortauthors{Monnier et al.}

\begin{document}


\title{Proper motions of new dust in the colliding-wind binary WR 140}


\author{J. D. Monnier\altaffilmark{1}, 
P. G. Tuthill\altaffilmark{2} and W. C. Danchi\altaffilmark{3}}

\altaffiltext{1}{Harvard-Smithsonian Center for Astrophysics, MS\#42,
60 Garden Street, Cambridge, MA, 02138}
\altaffiltext{2}{School of Physics, University of Sydney, NSW 2006, Australia}
\altaffiltext{3}{NASA Goddard Space Flight Center,
Infrared Astrophysics, Code 685, Greenbelt, MD 20771}
\email{jmonnier@cfa.harvard.edu, 
gekko@physics.usyd.edu.au, wcd@iri1.gsfc.nasa.gov}


\begin{abstract}
The eccentric WR+O binary system WR~140 produces dust for a few months
at intervals of 7.94~yrs coincident with periastron passage.  We
present the first resolved images of this dust shell, at binary phases
$\phi\sim0.039$ and $\sim0.055$, using aperture masking techniques on
the Keck-I telescope to achieve diffraction-limited resolution.
Proper motions of approximately 1.1~milliarcsecond per day were
detected, implying a distance $\simle$1.5\,kpc from the known wind
speed.  The dust plume observed is not as simple as the ``pinwheel''
nebulae seen around other WR colliding wind binaries, indicating the
orbital plane is highly inclined to our line-of-sight and/or the dust
formation is very clumpy.  Follow-up imaging in the mid-infrared and
with adaptive optics is urgently required to track the dust motion
further, necessary for unambiguously determining the orbital geometry
which we only partially constrain here.  With full knowledge of the
orbital elements, these infrared images can be used to reconstruct the
dust distribution along the colliding wind interface, providing a
unique tool for probing the post-shock physical conditions of violent
astrophysical flows.

\end{abstract}




\section{Introduction}
WR~140 is a prototypical colliding wind source consisting of a WC7
Wolf-Rayet and an O4-5 star, and has been extensively studied in the
radio, infrared, optical, ultra-violet, and even X-rays
\citep[e.g.,][]{moffat87,williams90,wb95,setia2001b}.  One of its more
remarkable properties is that dust forms over a period of a few months
close to periastron passage on each 7.94~yr orbit
\citep{hackwell76,williams78,williams90} when the two stars are only a
few AU apart.  Dust formation is apparently catalyzed at the wind-wind
interface in a layer of shock-compressed gas \citep{usov91}.  In this
paper we will use the orbital elements of \citet{williams90}, a 1985
periastron passage on JD~2446160 and period of 2900~days, for
estimating the orbital phases, but we hope refinement of these
parameters will be available soon based on data taken during the
recent periastron passage.

Recent advances in high-resolution imaging have allowed the dust
emission of other colliding wind sources to be resolved for the first
time. For WR~104 and WR~98a, continuous spiral plumes of dust were
observed, formed by colliding winds situated in approximately face-on
orbits with periods of $\sim$1~year \citep{tuthill99,monnier99}.  Dust
formation around the longer-period system WR~137 was observed to be
much more clumpy, but interpretation has been hampered by uncertain
orbital elements \citep{sergey1999}.  The recent periastron passage of
WR~140 in 2001 February afforded a unique opportunity to study this
prototype  by observing the formation and expansion of the
transient dust shell, in an attempt to both better understand the
3-dimensional orbital geometry of the binary and to probe exactly how
and where dust forms in these colliding wind systems.  This knowledge
is important for understanding the origins of interstellar dust
grains, the physics of strong astrophysical shocks, and the structure
of line-driven winds from Wolf-Rayets.

In this Letter, we report first results of our multi-wavelength
imaging program of WR~140.  We captured images of this dust shell
soon after periastron passage at two epochs, allowing the dust
morphologies and proper motions to be studied.
Our conclusions are preliminary and follow-up observations
are needed in order to more precisely interpret these new data.

\section{Observations}
\label{section:observations}

Aperture masking interferometry was performed by placing aluminum
masks in front of the Keck-I infrared secondary mirror.  This
technique converts the primary mirror into a VLA-style interferometric
array, allowing the Fourier amplitudes and closure phases for a range
of baselines to be recovered with minimal ``redundancy'' noise
\citep[e.g.,][]{baldwin86,jennison58}.  The Maximum Entropy Method
(MEM) \citep{sb84,mem86} has been used to reconstruct
diffraction-limited images from the interferometric data,
as implemented in the VLBMEM package by \citet{sivia87}.  In order to
check the reliability of the reconstructions, the MEM results have
been compared with those from the CLEAN reconstruction algorithm
\citep{hogbom74,cw81,pr84}.  Further engineering and performance
details may be found in \citet{pasp2000} and \citet{mythesis}.  In
addition, some short-exposure imaging was done without an aperture
mask (``Speckle Interferometry''), 
but the data was of insufficient quality to reconstruct reliable
images in most cases.

WR~140 was observed in 1999 July, 2001 June, and 2001 July at Keck-I
using the Near Infrared-Camera \citep{ms94,matthews96} in speckle
mode, using an integration time of 0.137\,s per frame. All important
observing information can be found in Table~\ref{table_nirc}.  This
table also contains photometric information on our multi-wavelength
observations, in particular the fraction of the flux in WR~140 from
the dust shell for each wavelength and for each epoch.  We also report
the magnitude difference between WR~140 and its point-source reference
star in each case.  Unfortunately, there are no published near-IR
magnitudes for the most of these calibrators, and so we are unable to
convert our relative photometry into absolute photometry at this time.
Assuming HD~192867 was constant between epochs, the 1.6\,\jmicron and 
2.2\,\jmicron fluxes of WR~140 did not significantly change between the
post-periastron epochs.

\begin{figure}[hbt]
\plotone{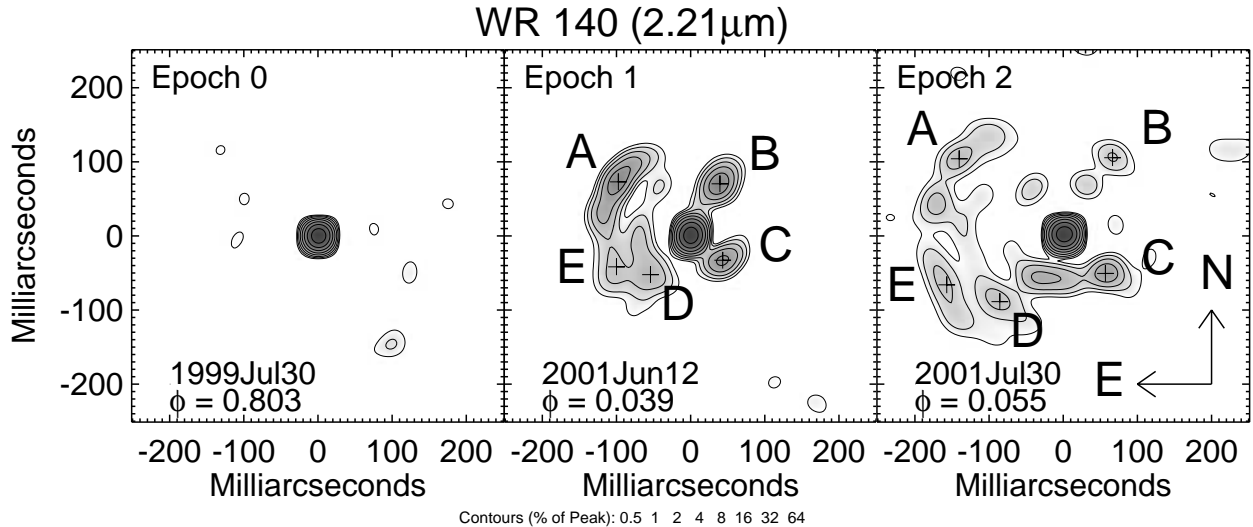}
\figcaption{2.2\jmicron image reconstructions of WR 140 at three epochs
using aperture masking.  The ``Epoch 0'' image was recorded prior to the
recent dust production episode, while the last two epochs show the expanding dust shell
after the binary periastron.
We have labeled five dominant dust
features as A through E.
Contour levels are 0.5, 1, 2, 4, 8, 16, 32, and 64\% of the peak.
\label{fig1alt}}
\end{figure}

Figure~\ref{fig1alt} shows MEM reconstructions of the WR~140 dust
shell at 3~epochs at 2.2\,\jmicron with $\sim$20~mas resolution.  The
Epoch~0 image was made well before periastron when we expected there
to be no circumstellar dust visible at this wavelength.  The Epoch~1
image was made by coadding 4 separate images made from independently
calibrated observations and has higher dynamic range than the image
from Epoch~2.  In addition, seeing conditions were worse in Epoch~2,
resulting in degraded imaging as evidenced by the higher noise level.

The test observation of WR~140 was taken in 1999 July
before the recent dust production episode, and there was no
evidence for any extended structure at 1.6\,\jmicron and 
2.2\,\jmicron as expected 
(see ``Epoch 0'' in Figure\,\ref{fig1alt}).  The
angular separation of the two components at that epoch could have been as
high as $\sim$14~mas based on \citet{williams90}, which would have been
marginally resolvable with Keck.  We inspected our data and found no
convincing evidence for a binary but can only set an upper limit of
$\sim$20~mas due to calibration uncertainties.  
We note that this binary should be easily resolvable with a long-baseline 
optical interferometer, and apastron in 2005 presents a wide enough target for
8-m class telescopes to resolve.

\begin{deluxetable}{lccccccc}
\footnotesize
\tablecaption{Observing Log and Relative Photometry for WR 140 at Keck-I
\label{table_nirc}}
\tablehead{
\colhead{Date} & \colhead{Aperture} & \colhead{ $\lambda_0$ } &
\colhead { FWHM } & \colhead{\# of} & & 
\colhead{$\Delta$Mag} & \colhead{\% Flux} \\
\colhead{(U.T.)}&  \colhead{Mask}& \colhead{ ($\mu$m) } & 
\colhead{ $\Delta\lambda$ ($\mu$m) } &\colhead{Frames} 
&\colhead{Calibrator}  & \colhead{(WR140-Cal)} & \colhead{From Dust}}
\startdata
1999 Jul 30 & None   & 1.6471 & 0.0176& 100 & HD~193631 & 0.14$\pm$0.03& $<$10 \\
            & None   & 2.2596 & 0.0531& 100 & HD~193631 &-0.01$\pm$0.03& $<$10\\
2001 Jun 12 & Annulus& 1.6575 & 0.333 & 300 & HD~192867 & 1.56$\pm$0.03& 43$\pm$7 \\
	    & Annulus& 2.2135 & 0.427 & 300 & HD~192867 & 0.75$\pm$0.03& 65$\pm$5\\
	    & Annulus& 2.269  & 0.155 & 100 & HD~193092 & 2.43$\pm$0.03& 68$\pm$5\\
	    & Annulus& 3.0825 & 0.1007& 100 & HD~192909 & 2.83$\pm$0.20& 81$\pm$15 \\
2001 Jul 30 & Annulus& 1.6575 & 0.333 & 100 & HD~192867 & 1.55$\pm$0.03& 32$\pm$7 \\
	    & Annulus& 2.2135 & 0.427 & 100 & HD~192867 & 0.72$\pm$0.03& 62$\pm$10\\
	    & Annulus& 3.0825 & 0.1007& 100 & HD~193092 & 1.55$\pm$0.03& 92$\pm$5 \\
	    & None   & 1.6471 & 0.0176& 100 & HD~192867 & 1.68$\pm$0.05& 34$\pm$10\\
            & None   & 2.2596 & 0.0531& 100 & HD~192867 & 0.89$\pm$0.04& 66$\pm$10\\
	    & None   & 3.0825 & 0.1007& 100 & HD~192867 &-0.06$\pm$0.03& 79$\pm$10\\
\enddata
\end{deluxetable}

\begin{deluxetable}{ccccccl}
\tablecaption{Relative Positions of WR 140 Dust Features
\label{table_astrometry}}
\tablehead{
 & \multicolumn{2}{c}{2001 Jun 12} &\multicolumn{2}{c}{2001 Jul 30} & 
\colhead{Velocity} & \colhead{Date} \\
\colhead{Feature} & \colhead{R\tablenotemark{(a)}~~(mas)} & \colhead{PA\tablenotemark{(b)}~~(\arcdeg)} &
  \colhead{R (mas)} & \colhead{PA (\arcdeg)} & \colhead{(mas/day)} & 
\colhead{Ejected\tablenotemark{c}}}
\startdata
A & 122 & 54 & 175 & 54 & 1.10$\pm$0.20 & 2001 Feb 21$\pm$20 \\
B & 79  & 331& 123 & 328& 0.92$\pm$0.20 & 2001 Mar 17$\pm$18 \\
C & 54  & 231& 76  & 227& 0.46$\pm$0.20 & 2001 Feb 14$\pm$52\\
D & 77  & 134& 125 & 136& 1.08$\pm$0.40 & 2001 Mar 27$\pm$28 \\
E & 110 & 113& 172 & 113 &1.29$\pm$0.40 & 2001 Mar 30$\pm$28 \\
\enddata
\tablenotetext{a}{R is the distance of feature from central source of
WR~140 in milliarcseconds.}
\tablenotetext{b}{PA is the position angle of the feature in degrees East of North.}
\tablenotetext{c}{Errors in table are in units of days.}
\end{deluxetable}

\section{Discussion}

Tracks of the motion of five dust features, labelled A through E on
Figure\,\ref{fig1alt}, are compiled in Table\,\ref{table_astrometry}.
The arc of dust to the north-east of the central source (feature A)
appears to be moving at an angular velocity of
$1.1\pm0.2$\,milli-arcseconds per day, similar to the velocity of the
eastern features D and E.  The western features B and C appear to have
slower velocities, perhaps indicating they are moving more out of the
plane of the sky.  We caution that image reconstructions of weak
features from interferometry data can suffer from systematic errors,
including apparently clumpy features in place of smooth faint
nebulosity.  The apparent position angles of each marked feature
remained unchanged between epochs (to within errors of a few degrees),
giving confidence that we are indeed tracking physical radial motion
of dust and are not strongly affected by image fidelity problems.
The positions of features of D and E are most uncertain, since the local
maxima in Epoch 1 are relatively weak and could be significantly influenced by
noise.  We have estimated the relative position uncertainty to be $\sim$10~mas 
for Features A-C, and $\sim$20~mas error for Features D \& E, and this is 
reflected in the velocity error compiled in the table.

If we assume that feature A is moving at the WR wind speed
\citep[$\sim$2860km/s;][]{williams90} and in the plane of the sky,
then we set a distance upper limit to WR 140 of
$\sim$1.5\,kpc, which is close to previous estimates based on the
apparent luminosity of the O4-5 companion \citep[1.3~kpc for main
sequence, 1.7~kpc for supergiant]{williams90}.  The uncertainty of
20\% on the dust feature velocities can be substantially reduced
with continued monitoring using adaptive optics or high SNR mid-IR
measurements on an 8-m class telescope.

Estimated times for ejection of dust features from the
binary (assuming constant velocity and extrapolating back to the
central stars) can be found in Table\,\ref{table_astrometry}.
Dates are consistent with the periastron prediction of
\citet{williams90}: 2001 February 19.  This is further
borne out by the fact that dust in the partial arc of Feature~A is
roughly equidistant from the star, as expected if formed over a short
period of time as the O-star swung quickly around the WR star just at
periastron.  If this dust arc was formed in this manner, this would imply that
the O-star was located roughly
East of the WR star on the sky at periastron.
However, it is surprising that Features B and C seem to
have also formed close to this time, considering they are moving in very
different directions from feature A.

It is tempting to interpret the eastern arc of dust in
Figure\,\ref{fig1alt} (comprising features A, E and D) as part of an outward 
spiraling clockwise plume similar to those seen around pinwheel nebulae,
implying that the underlying binary has an counter-clockwise orbit.
This orbital motion is difficult to reconcile with preliminary VLBA observations of
the non-thermal radio emission by Beasley (2001, private communication)
which indicate a clockwise orbit.  Furthermore, a simple pinwheel
nebula type model (as applicable for WR~104 and WR~98a) can not easily
explain the additional emission to the northwest and southwest (features B and C).
We must look to a more complex model to understand these observations.

Naively, we expect the dust to form in a thin cone-like layer between
the colliding winds, which gets twisted into a spiral with the orbit
of the underlying binary \citep{tuthill99}. 
This cone forms because the winds have unequal momenta and the 
geometry of the interface region asymptotically resembles a cone bending
away from the star with the stronger wind \citep{usov91,crw96}.
For WR~140, the cone (full)
opening angle can be estimated to be $\sim$74\arcdeg~~by 
using estimated wind velocities and densities in  \citet{williams90} and
the analytical formulation of \citet{eu93}.
Regarding the orientation of the orbit, 
we have multiple reasons to think that the inclination of the
orbit is large ($\sim$60\arcdeg): from time-variable shadowing effects in the radio
\citep{wb95} and radial velocity measurements combined with
the expected masses of the components \citep{williams90}.

\begin{figure}
\plotone{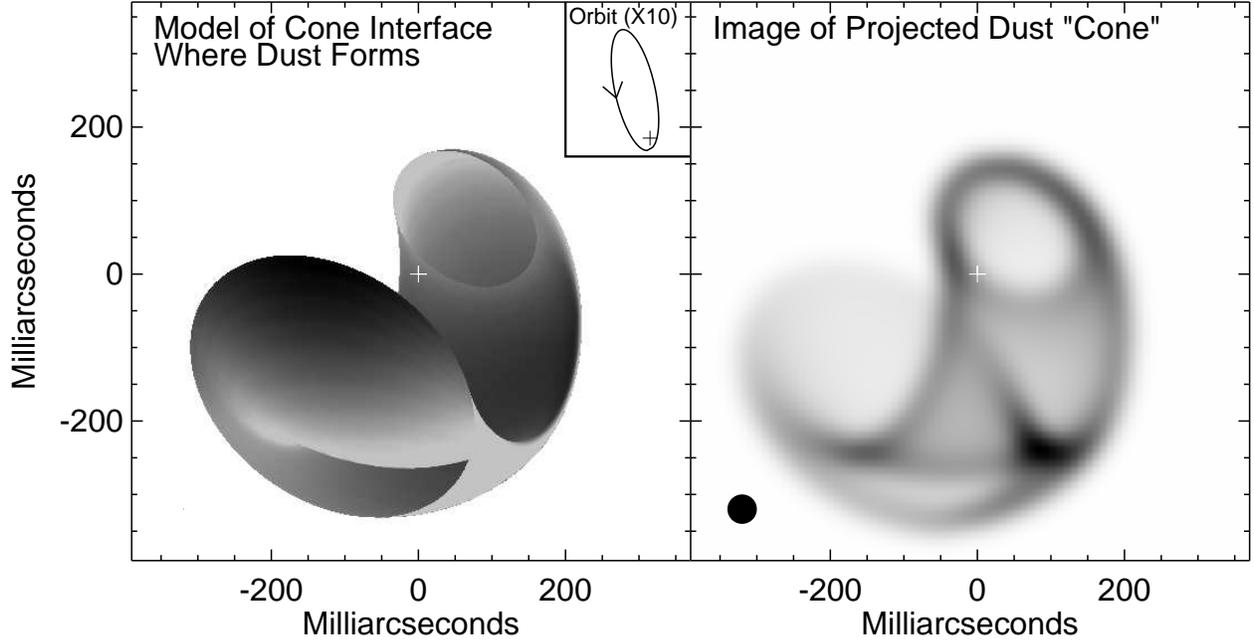} \figcaption{This figure illustrates the
difficulty in interpreting the dust shell images of WR~140 without
strict {\em a priori} constraints on the full (3-dimensional) orbital
elements.  The left panel shows the wind-wind interface region where
dust is expected to form in an colliding wind eccentric binary like WR~140 at 60\arcdeg\
inclination, assuming ``triggered'' dust formation around periastron
(see text for details).  The plus signs mark the location of the
binary system, and a 10$\times$ expanded view of the binary orbit of
the O-star around the WR star is shown in the inset plot.  The right
panel shows a projection of this dust cone onto a plane, smoothed to
40~mas resolution and plotted with a linear scale.  Features in this
image should correspond to a real observed image if the dust emission
is optically thin, although this particular model was {\em not fit} to
the observed data.  Arcs and filaments are present due to
limb-brightening effects.
\label{fig_mess}}
\end{figure}

To understand the difficulty in interpreting the
near-infrared morphology under such circumstances, consider
Figure~\ref{fig_mess}.  This figure shows a portion of the cone
interface for a colliding wind binary system with the known (spectroscopic) 
orbital elements of WR~140
\citep{williams90}, an inclination angle 60\arcdeg, and an observation
date about 0.5~yr after periastron.  Based on previous findings for
WR~140, this model also assumes that dust is only formed during a
``trigger''-zone of the orbit, within 0.15~years of periastron
passage (before and after), 
and the conic surface between the two winds is only rendered for this span.  This
figure also displays what this model might look like on the sky if the
dust was optically thin and had constant emissivity; that is, the
``cone of dust'' has been projected onto the plane of the sky.  Arcs, filaments and
complex shapes are formed in this manner due, depending on the orbital
orientation, cone opening angle, and the span of the trigger-zone.

For high inclination binaries such as WR~140, arcs of emission might be tracing the
limb-brightening of the dust cone with our line-of-sight, giving us a
sensitive probe of the dust distribution if the orbit geometry is
known from other means. 
With full orbital elements, hopefully
derivable from the VLBA monitoring program or mid-infrared imaging, we
expect to be able to invert our images of the dust in order to map out
the dust density along the wind-wind interface cone.  However without
independent constraints on the orbital inclination, trigger zone, and other
parameters, we are unable to disentangle orbital effects from
uncertainties in the dust density distribution.

\begin{figure}
\plotone{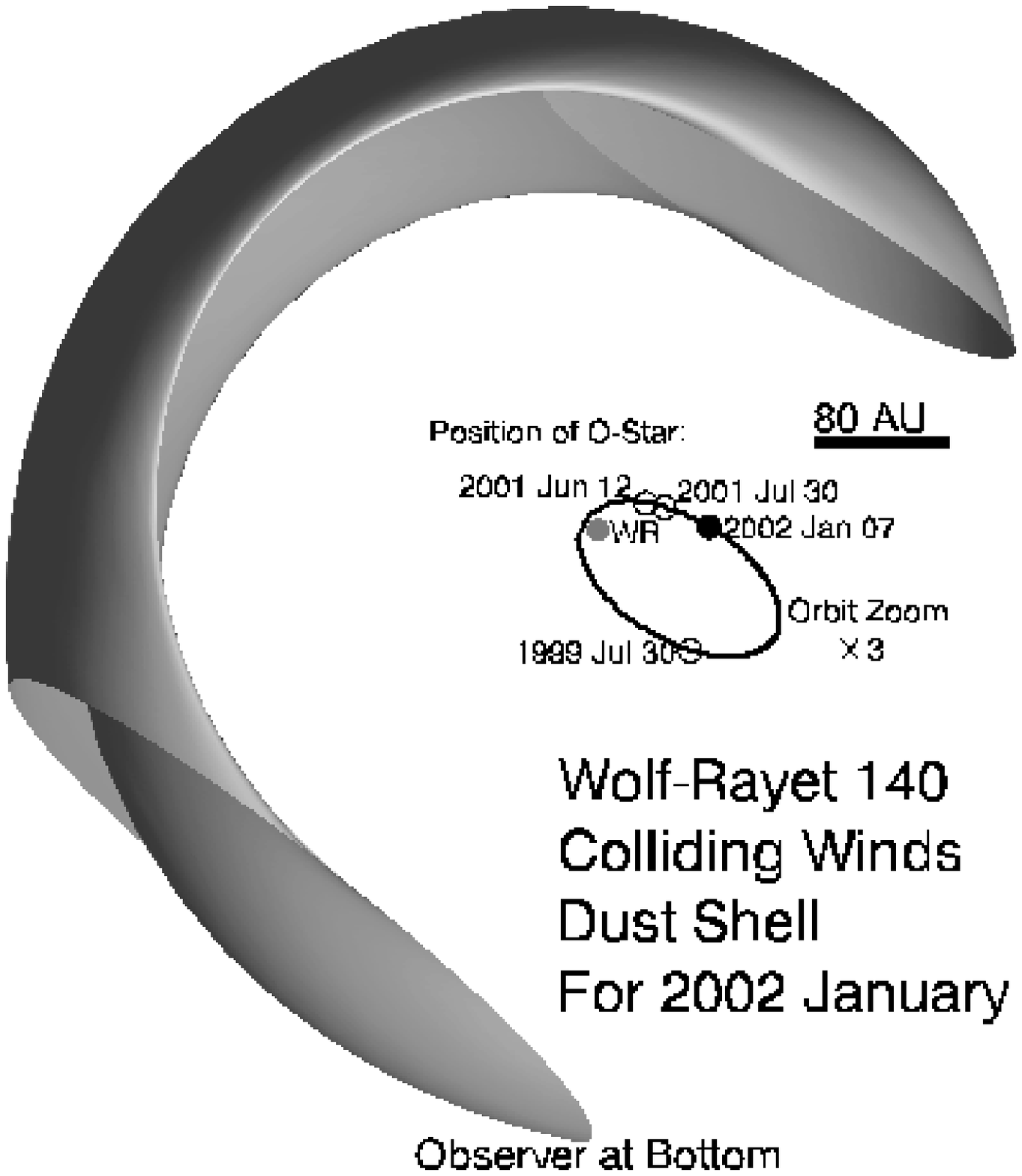} \figcaption{This figure shows our best
estimate of the observing geometry for WR~140 based on the 
\citet{williams90} orbital parameters and these new observations.
The location of the O-star relative to the WR is marked for each 
imaging epoch, and
the observer is located toward the bottom at some 
poorly-determined inclination.   The orbit has been magnified by a factor
of 3 compared to the dust plume.
\label{fig_extra}}
\end{figure}

Figure~\ref{fig_extra} shows a schematic of the WR~140 
binary orbit and dust shell which is qualitatively consistent with our new
results and the orbital elements of \citet{williams90}.  
The triggered dust shell is rendered from a top-down perspective with
the observer located at the bottom on the image, viewing the dust from
some currently unknown inclination angle.  This basic picture has the correct
orbital rotation as implied by the preliminary VLBA results, and can explain why
the projected velocities of the eastern dust features are fast compared to
the slower-moving western features that have significant out-of-the-plane 
motions.

WR 137 is another eccentric WR+O system which episodically produces
dust. \citet{sergey1999} found evidence for production of a dust
``jet'' in one direction and dust clouds moving away in the opposite
direction around periastron.  These authors hypothesized that dust
production might occur in localized clumps and not smoothly throughout
the wind collision zone.  The dust features in the circumstellar
environment of WR~140 could be interpreted also as isolated dust
clumps, and we require additional orbital information to investigate this
further.  Such clumps might form through cooling instabilities in the
post-shock gas or could be seeded 
by dense clumps in the WR wind itself \citep[e.g.,][]{mr94}.

Lastly, our interpretation of the WR~140 dust shell is somewhat
hindered by the fact that we are not sure if the near-infrared dust
opacity is optically thick, although our observations are long enough
after periastron that the dust clouds have likely significantly expanded and 
become optically thinner.
Dust emission in the mid-infrared should surely be optically-thin and thus
trace out the full volume of the emitted dust.  
As the dust cools rapidly moving away from the central sources 
\citep[as carefully shown in][]{williams90}, further
near-infrared imaging of the thermal emission 
will be difficult until the next periastron in 2009.
However mid-infrared emission is observable for longer than near-infrared,
and also scattered light imaging in the near-infrared 
might be possible using adaptive optics.

\section{Conclusions}
We have presented the first resolved images of the transient dust
shell formed around the WR 140 colliding wind binary once every 7.94
years during periastron passage.  Our multi-epoch images show clear
expansion of the nebula, but resist detailed interpretation due to
inadequate knowledge of the orbital geometry. We interpret the
emission as indicating a highly inclined view of the dust plume and/or
very clumpy and inhomogeneous dust production.
We have shown that an optically-thin ``cone'' of dust generated in a WR~140-like 
system can indeed show a variety
of complex features if viewed near edge-on.
With full orbital
elements, we may be able to use these images to map out the dust
distribution on the surface of the wind-wind interface cone between
the colliding winds. Although the near-infrared thermal emission quickly
fades, the dust shell will be readily observable in the mid-infrared
for years and should be well-resolved from the central source using
8-m class telescopes.  We urge the community to take advantage of the
unique opportunity to observe this rapidly evolving system for the
next few years in order to extract maximal information on this
important prototypical colliding wind source.

\acknowledgments
{We wish to thank P. Williams and C. H. Townes for their support of this 
experiment, and B. Schaeffer for last minute scheduling changes at
Keck that were critical.
JDM acknowledges support
from a Center for Astrophysics Fellowship at the Harvard-Smithsonian
Center for Astrophysics.
This research has made use of the SIMBAD
database, operated at CDS, Strasbourg, France, and NASA's Astrophysics
Data System Abstract Service.  The data presented herein were
obtained at the W.M. Keck Observatory, which is operated as a
scientific partnership among the California Institute of Technology,
the University of California and the National Aeronautics and Space
Administration.  The Observatory was made possible by the generous
financial support of the W.M. Keck Foundation.
}
\bibliographystyle{apj}
\bibliography{apj-jour,WR140,Thesis}


\clearpage

\end{document}